\documentclass[11pt]{article}
\usepackage{moriond,epsfig}

\newcommand{\GeV}{\mbox{GeV}}

\newcommand{\spom} {\mbox{$\scriptstyle \mathrm{I}\! \mathrm{P}$}}

\def\lsim{\mathrel{\rlap{\lower4pt\hbox{\hskip1pt$\sim$}}
    \raise1pt\hbox{$<$}}}         
\def\gsim{\mathrel{\rlap{\lower4pt\hbox{\hskip1pt$\sim$}}
    \raise1pt\hbox{$>$}}}         

\def\GeV{\mbox{\rm GeV}\/}
\def\GeV2{\mbox{\rm GeV$\,^2$}}

\def\Q2{\mbox{$Q^2$}}
\def\F2{\mbox{$F_2$}}

\def\be{\begin{equation}}
\def\ee{\end{equation}}
\def\bea{\begin{eqnarray}}
\def\eea{\end{eqnarray}}

\begin{document}
\vspace*{4cm}
\title{INCLUSIVE DIFFRACTION AT HERA}

\author{ M. RUSPA (on behalf of the H1 and ZEUS Collaborations) }

\address{Dipartimento di Fisica Sperimentale, via P.Giuria 1, I-10025 TORINO, Italy \\
E-mail: ruspa@to.infn.it}

\maketitle\abstracts{The diffractive dissociation of virtual photons, 
$\gamma^*p \rightarrow Xp$, has been studied in $ep$ interactions at 
HERA with the H1 and ZEUS detectors. The data are presented in terms 
of the diffractive structure function $F_2^D$ 
and of the diffractive cross sections $d\sigma/dt$ and $d\sigma/dM_X$. 
The $t$-dependence of the cross section is measured. 
The Pomeron intercepts extracted from diffractive and inclusive $ep$ 
interactions are compared. The result is further interpreted studying the 
dependence of the ratio between the diffractive to the 
inclusive cross section on the photon-proton center of mass energy. 
Recent data on the $Q^2$ dependence of the diffractive cross section at 
$0.17 < Q^2 < 0.70$ GeV$^2$ constrain the transition in diffraction 
from the perturbative high $Q^2$ region to the photoproduction limit.   
}

\section{Diffraction at HERA}

\begin{figure}[bh]
\begin{center}
    \mbox{\epsfig{file=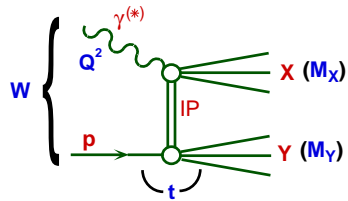,bbllx=116pt,bblly=180pt,
bburx=509pt,bbury=407pt, width=4.7cm, clip=}}
\end{center}
\caption{The generic HERA diffractive process of the type $\gamma^*p\rightarrow XY$.
\label{fig:diagram}}
\end{figure}

The diffractive dissociation of real photons, $\gamma p \rightarrow Xp$, 
can be treated in complete analogy with hadron-hadron diffractive 
interactions and therefore can be described by Regge phenomenology. 
Whithin this framework, diffractive 
reactions at high energy are dominated by the exchange of a trajectory with 
the quantum numbers of the vacuum, referred to as the ``Pomeron'' trajectory.  
By contrast HERA allows the investigation of the partonic nature of diffraction in 
deep-inelastic scattering (DIS) using virtual photons as probes. The most 
general HERA diffractive DIS process is shown in Fig.~\ref{fig:diagram}. 

A photon of virtuality $Q^2$ (coupled to the electron) interacts with a 
proton at a center of mass energy $W$ and squared four momentum transfer $t$ 
to produce two distinct hadronic systems of masses $M_X$ and $M_Y$, 
respectively. 
The system $X$ is usually reconstructed in the central detector and is well separated in rapidity from the proton remnant system Y which, for small masses $M_Y$, escapes through the forward beam hole of the detector. The photon dissociative processes, where the proton remains intact ($Y=p$), are selected at HERA either by requiring an absence of particles in the forward region of the central detector (``rapidity gap method'') or by measuring the scattered proton in special downstream detectors, the H1 Forward Proton Spectrometer (FPS) and the ZEUS Leading Proton Spectrometer (LPS).  The former method allows large statistics, but contains a background of double-dissociative events, $ep \rightarrow eXY$. Conversely, the latter method selects events $ep \rightarrow eXp$, but is statistically limited because of the small acceptance of the forward spectrometers. 

\subsection{Diffractive cross section and structure function}\label{subsec:dcssf}

The diffractive cross section $\sigma ^{diff}_{ep}$ for the process $ep \rightarrow eXp$ is related to the cross section $\sigma ^{diff}_{\gamma^*p}$ for the diffractive dissociation of virtual photons, $\gamma ^*p \rightarrow Xp$:
\begin{equation}
\frac{d\sigma_{\gamma^*p}^{diff}} {dM_X}=\frac {2\pi Q^2y} {\alpha(1+(1-y)^2)}
\cdot \frac{d^3\sigma_{ep}^{diff}}{dQ^2dydM_X}.
\label{eq:sigma}
\end{equation}

The diffractive cross section $\sigma ^{diff}_{ep}$ can also be also be expressed in terms of the diffractive structure function $F_2^{D(4)}(\beta, Q^2, X_{\spom}, t)$, where $x_{\spom}$ may be interpreted as the fraction of the proton longitudinal momentum carried by the Pomeron and transferred to the $X$ system and $\beta$ as the fraction of the Pomeron momentum carried by the quark coupling to the photon:  
\begin{equation}
F_2^{D(4)}(\beta,Q^2,x\spom,t)=\frac{\beta Q^4}{4\pi \alpha^2 (1-y+y^2/2)} \cdot 
\frac{d\sigma_{\gamma^*p}^{diff}}{d\beta dQ^2 dx_{\spom} dt}.
\label{eq:func}
\end{equation}

\noindent
$\alpha$ is the electromagnetic coupling costant and the contribution of the longitudinal structure function and the effect of the $Z^0$ exchange have been neglected. With this definition, $F_2^{D(4)}$ has dimension of GeV$^{-2}$. Integration over $t$ yields the three-fold diffractive structure function $F_2^{D(3)}(\beta, x_{\spom}, Q^2)$.

\section{Results}

\subsection{t-dependence}\label{subsec:tdep}
\begin{figure}[p]
\begin{center}
    \mbox{\epsfig{file=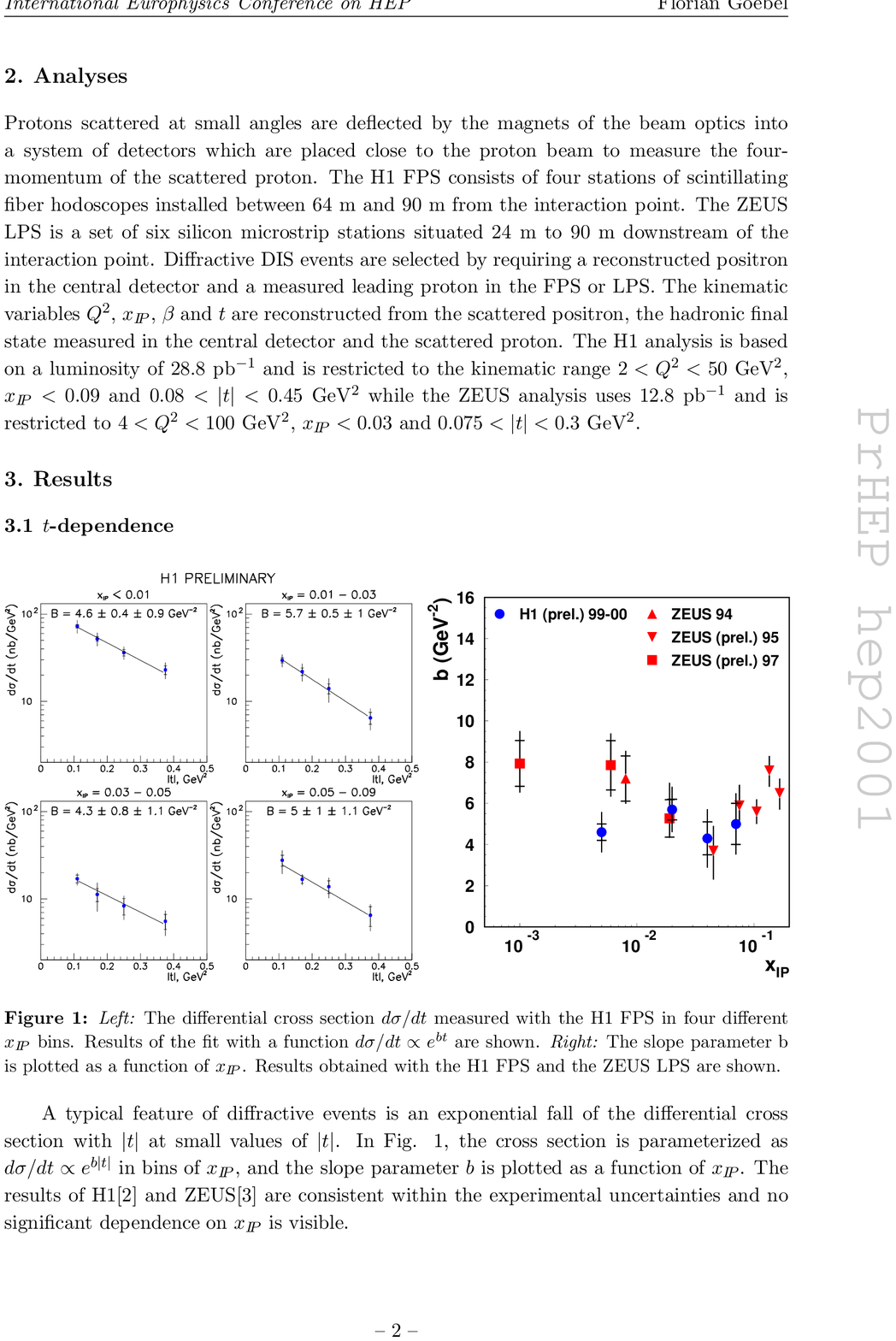,bbllx=52pt,bblly=251pt,
bburx=540pt,bbury=502pt,width=16cm,clip=}}
\end{center}
\caption{{\it Left:} The differential cross section $d\sigma/dt$ measured with the H1 FPS in four different $x_{\spom}$ bins. The exponential fit $e^{b|t|}$ is also shown. {\it Right:} the slope parameter $b$ is plotted as a function of $x_{\spom}$. Results obtained with the H1 FPS and the ZEUS LPS are shown.~~~~~~~~~~~~~~~~~~~~~~~~~~~~~~~~~~~~~~~~~~~~~~~~~~~~~~~~~~~~~~~~~~~~~~~~~~~~~~~~~~  
\label{fig:t}}
\end{figure}

A typical features of diffractive events is an exponential fall of the differential cross section with $|t|$ at small values of $|t|$. In Fig.~\ref{fig:t} the cross section is parametrized as $d\sigma/dt \propto e^{b|t|}$ in bins of $x_{\spom}$ and the slope parameter $b$ is plotted as a function of $x_{\spom}$. 

\subsection{$x_{\spom}$ dependence of the diffractive structure function}\label{subsec:xddsf}

\begin{figure}[htbp]
\hspace*{-1.8cm}
\begin{minipage}[p]{1.085\linewidth}
  \begin{center}
    \leavevmode  
    \mbox{\epsfig{file=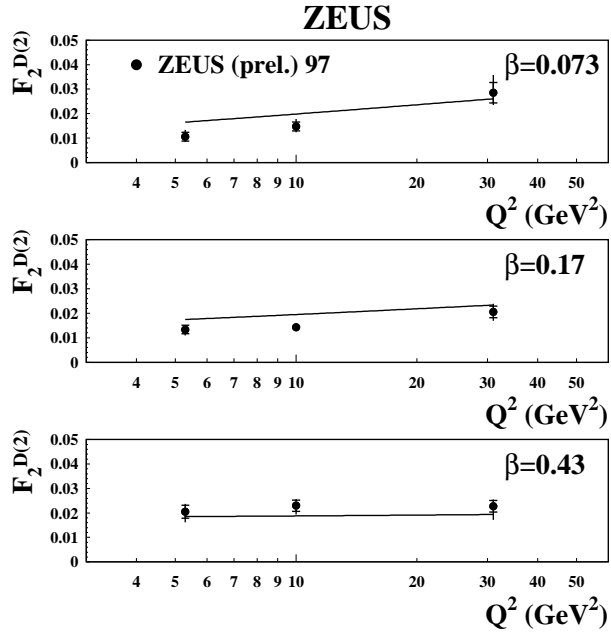,width=9.cm}}
  \end{center}
\end{minipage}\hfill
\caption{{\it Left}: The diffractive structure function $x_{\spom}F_2^{D(3)}$ measured with the FPS is plotted as a function of $x_{\spom}$ in bins of $\beta$ and $Q^2$. A comparison with the H1 rapidity gap measurement is also shown. {\it Right}: The values of $F_2^{D(2)}(\beta,Q^2)$ obtained by ZEUS from a fit to $F_2^{D(3)}$ are shown as a function of $Q^2$.~~~~~~~~~~~~~~~~~~~~~~~~~~~~~~~~~~~~~~~~~~~~~~~~~~~~~}
\label{fig:f2d3}
\end{figure}

In Fig.~\ref{fig:f2d3} the values of $x_{\spom}F_2^{D(3)}$ obtained using H1 FPS~\cite{h1:t} are compared with the recent H1 measurement~\cite{h1:gap} based on the rapidity gap method. The data are extrapolated to the full $t$ range using the measured $t$ dependence. Within uncertainties, the leading proton data are in good agreement with the rapidity gap data, indicating that the contribution of proton dissociation to the rapidity gap analysis is small.
In Fig.~\ref{fig:f2d3} $F_2^{D(2)}$ measured using ZEUS LPS~\cite{zeus:t} is also shown as a function of $\beta$ and $Q^2$. Clear scaling violation are seen in the low $\beta$ bins.

\subsection{W dependence of the diffractive cross section}

\begin{figure}
\begin{center}
\hspace{-1cm}
    \mbox{\epsfig{file=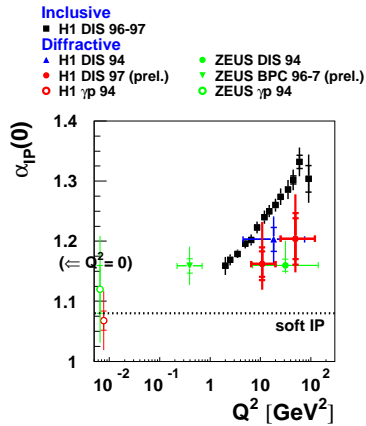,bbllx=0pt,bblly=12pt,
bburx=256pt,bbury=347pt, width=5cm, clip=}}
\end{center}
\caption{$\alpha_{\spom}(0)$ extracted from the energy dependence of selected inclusive and diffractive HERA data.
\label{fig:Pomeron}}
\end{figure}

The study of the $x_{\spom}$ dependence of $F_2^D$ is equivalent to that of the $W$ dependence of $d\sigma^{diff}/dM_X$.
Interpreting the $x_{\spom}$ dependence of the data at fixed $\beta$ and $Q^2$ in a Regge motivated model, whereby $x_{\spom}F_2^D \sim A(\beta, Q^2)~~(1/x_{\spom})^{2\bar{\alpha_{\spom}(t)-1}}$, one can extract the $t$ averaged value of the effective exchanged trajectory 
	$\bar{\alpha}_{\spom} =
	\alpha_{\spom}(0) -
	\alpha_{\spom}'\cdot |\bar{t}|$.
Equivalently , according to Regge formalism, the $W$ dependence of the diffractive cross section can be parametrized as $d\sigma^{diff}/dM_X \propto (W^2)^{2\bar{\alpha}_{\spom}-2}$.
Different values of the effective Pomeron trajectory extracted from both diffractive~\cite{h1,trova1,trova2,kowa} and inclusive~\cite{trova3} $ep$ measurements are presented in Fig.~\ref{fig:Pomeron}. 

\begin{figure}[phb]
\begin{center}
    \mbox{\epsfig{file=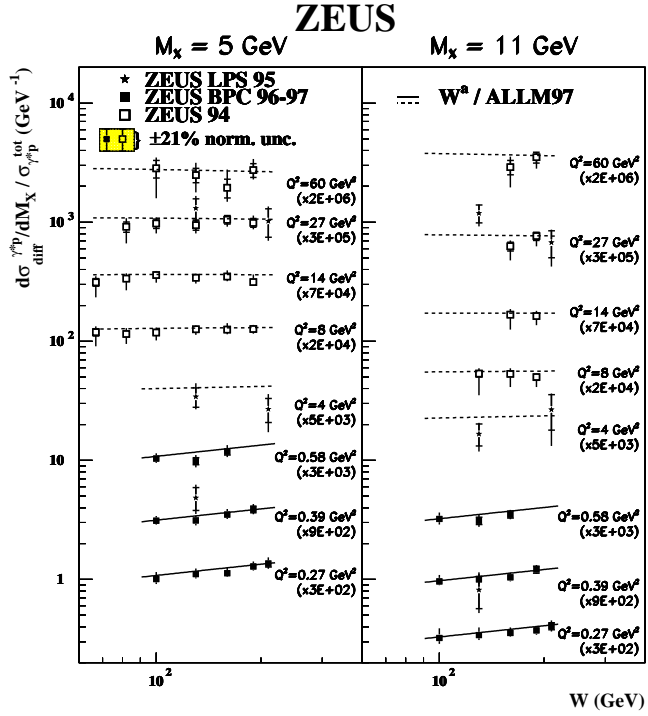,bbllx=0pt,bblly=12pt,
bburx=579pt,bbury=634pt, width=9cm, clip=}}
\end{center}
\caption{The values of the ratio $d\sigma^{diff}/dM_X/\sigma^{tot}_{\gamma^*p}$ for different $W$ and $M_X$ bins as a function of $Q^2$.
\label{fig:w}}
\end{figure}

At large $Q^2$, the Pomeron intercept describing the inclusive data is larger than that from the diffractive data, while it has been observed~\cite{antonio} that in photoproduction and low $Q^2$ electroproduction the effective intercept in the diffractive and inclusive case are compatible. In fact Fig.~\ref{fig:w}, where the ratio of the diffractive to the total cross section is plotted for $Q^2$ and $M_X$ bins as a function of $W$, indicates that the energy dependences of the inclusive and diffractive cross sections are rather similar at large $Q^2$, in contrast to the expectation from Regge theory. However, the data for $Q^2<1$ GeV$^2$ show a rise in $W$, consistent with Regge theory.  

\subsection{$Q^2$ dependence of the diffractive cross section}
\begin{figure}[ph]
\begin{center}
    \mbox{\epsfig{file=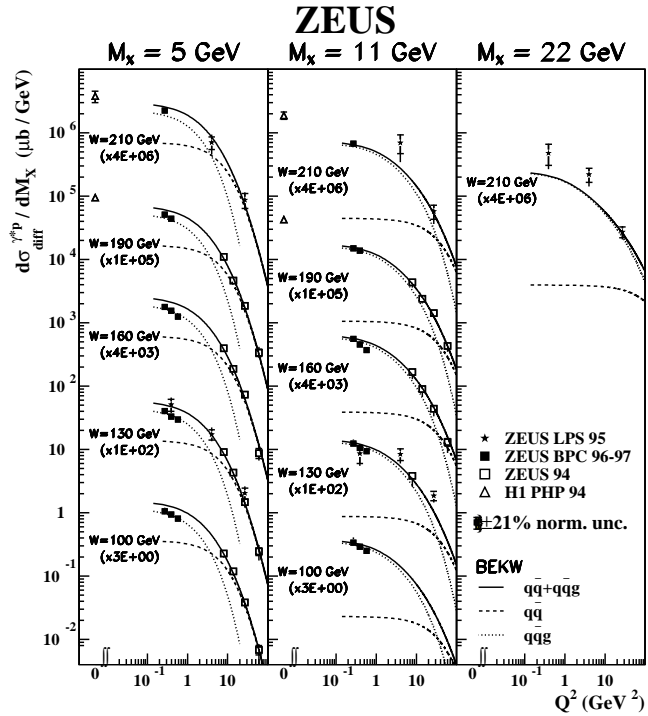,bbllx=0pt,bblly=12pt,
bburx=579pt,bbury=634pt, width=9cm, clip=}}
\end{center}
\caption{The values of $d\sigma_{\gamma^*p}^{diff}/dM_X$ for different $W$ and $M_X$ bins as a function of $Q^2$. The lines are the result of the BEKW parametrization described in the text.~~~~~~~~~~~~~~~~~~~~~~~~~~~~~~~~~~~~~~~~~~~~~~~~~~~~~~~~~~~~~~~~~~~~~~~~~~~~~~~~~~~~~ 
\label{fig:q}}
\end{figure}

Fig.~\ref{fig:q} shows the diffractive cross section, 
$d\sigma_{\gamma^*p}^{diff}/dM_X$, 
as a function of $Q^2$ in different bins of $M_X$ and $W$. The recent ZEUS measurement~\cite{antonio} at low $Q^2$ is shown toghether with previous ZEUS~\cite{kowa} and H1~\cite{h1} measurements respectively in the DIS region and in photoproduction. A sharp change of the $Q^2$ dependence of $d\sigma^{diff}/dM_X$ is apparent with decreasing $Q^2$, reminiscent of the behaviour of the photon-proton cross section, $\sigma^{\gamma^*p}_{tot}$, which also exhibits~\cite{total} a flattening of the $Q^2$ dependence for  $Q^2 \lsim 1$ GeV$^2$. This is consistent with the expectation, based on the conservation of the electromagnetic current, that $\sigma^{tot}_{\gamma^*p}$ $\rightarrow$ constant or, equivalently, that $F_2 \propto Q^2$ as $Q^2 \rightarrow 0$. Moreover Fig.~\ref{fig:q} shows that the main features of the data are reproduced by a parametrization based on the model of Bartels {\it et al.} (BEKW)~\cite{bekw}, in which the dominant contributions to the diffractive structure function come from the fluctuations of the photon into either a $q\bar{q}$ pair or a $q\bar{q}g$ pair, being the first component dominant at large $\beta$ and the second one dominant at small $\beta$. As $Q^2$ decreases, for a fixed value of $M_X$ $\beta$ also decreases: the data indicate the increasing importance of the contribution of $q\bar{q}g$ production at low $Q^2$. 

\end{document}